\documentclass[a4paper,11pt]{article}
\pdfoutput=1 

\usepackage{jheppub} 

\usepackage{multirow}
\newcommand{\ba}{\begin{eqnarray}}
\newcommand{\ea}{\end{eqnarray}}

\DeclareMathOperator{\csch}{csch}

\usepackage[T1]{fontenc} 

\title{\boldmath Warm inflation in general scalar-tensor theory of gravity}

\author[a]{Waluka Amaek,}
\author[a,b]{Apirak Payaka,}
\author[a,b]{Phongpichit Channuie}

\affiliation[a]{College of Graduate Studies, Walailak University, Thasala, Nakhon Si Thammarat, \\80160, Thailand}
\affiliation[b]{School of Science, Walailak University, Thasala, Nakhon Si Thammarat, \\80160, Thailand}

\emailAdd{waluka.am@gmail.com, apirak.pa@mail.wu.ac.th, channuie@gmail.com}

\abstract{In this work, we investigate warm inflationary models in the context of a general scalar-tensor theory of gravity which is coupled to radiation through a dissipation term. We first derive the potential of exponential and hyperbolic tangent forms. We consider a dissipation parameter of the form $\Gamma = C_{1}T$ with $C_1$ being a coupling parameter and focus only on the strong regime of which the interaction between inflaton and radiation fluid has been taken into account. We compute inflationary observables and constrain the parameters of our model using current Planck 2018 data. From our analysis, we discover that the weak coupling limit $\xi\ll 1$ is needed in order to have the derived $n_s$ and $r$ consistent with the Planck 2018 observational constraints. Particularly, we constrain the potential scale $V_{0}$ of the models.}

\begin{document} 
\maketitle
\flushbottom

\section{Introduction}
Since its first proposal, an inflationary scenario is a well-established paradigm describing an
early universe and  becomes a pillar of modern cosmology. In other words, it was used to describe a broad range of observed phenomena including the anisotropy of the
cosmic microwave background (CMB) consisting of the small temperature fluctuations in the blackbody
radiation left over from the Big Bang and a mechanism for generating the primordial energy density
perturbations seeding for a late time large scale structure. In the standard picture, the interactions between the inflaton with other fields resulting the (partial) decay of the
inflaton into ordinary matter and radiation is needed. This was so-called \lq\lq cold inflation\rq\rq \cite{Starobinsky:1980te,Sato:1980yn,Guth:1980zm,Linde:1981mu,Albrecht:1982wi}. The decay process however plays an important role
at the end of the slow-roll mechanism leading to the standard “(p)reheating” paradigm, see e.g. \cite{Linde:2005ht,Albrecht:1982mp,Abbott:1982hn}.

However, an alternative approach of cold inflation may be possible. If one introduces a coupling between
inflaton and radiation, the energy density of radiation can be maintained almost a constant during inflation and the (p)reheating is unnecessary This alternative scenario was known as \lq\lq warm inflation\rq\rq and deserves some major attention \cite{Berera:1995wh,Berera:1996fm,Berera:1999ws,Taylor:2000ze,Hall:2003zp,Berera:2008ar,Bartrum:2013fia}. To generate the thermal bath in the standard cosmology, this warm inflation scenario has received much attention. To be more precise, it was originally proposed to resolve some problems in the standard cold inflation picture [1, 2], for instances, providing sufficiently hot thermal bath.

Warm inflation has been recently studied in many different theories. For instance, a possible
realization of warm inflation owing to a inflaton field self-interaction was conducted in Ref.\cite{Dymnikova:2000gnk}. Additionally, a number of investigations of minimal  and  non-minimal  coupling  to  gravity were investigated in Refs.\cite{Panotopoulos:2015qwa,Benetti:2016jhf,Motaharfar:2018mni,Graef:2018ulg,Arya:2018sgw,Kamali:2018ylz}. Very recently, the authors of Refs.\cite{Samart:2021eph,Samart:2021hgt} investigated the Higgs-Starobinsky (HS) model as well as a non-minimally coupled scenario with quantum-corrected self-interacting potential in the context of warm inflation.

In this work, we investigate warm inflationary models in the context of a general scalar-tensor theory of gravity. We will demonstrate that our results by introducing a coupling between inflaton and radiation -- a dissipative term, can complete the radiation dominated Universe at the end of inflation and confront them with the last Planck satellite data.

The paper is organized in the following way. In Section \ref{s2}, we will take a short recap of the formalism in the general scalar-tensor theory and present our derivations on the potentials of the exponential (E) and hyperbolic tangent (T) forms. All relevant dynamical equations in the non-minimal coupling warm inflation under the slow-roll approximation are determined in Section \ref{s3}. Here the
spectral index and the tensor-to-scalar ration will be derived. In Section \ref{s4}, we will compare the results in this work with the observational data. Finally, we close this paper by providing discussions and conclusions in the last section.

\section{General scalar-tensor theory: a short review}\label{s2}
The action of a general scalar-tensor theory in the Jordan frame takes the form
\begin{eqnarray}
S_J = \int d^4x\sqrt{-g}\,\bigg[ -\frac{1}{2}\Omega^{2}(\phi)\,M_p^2\, R + \frac{1}{2}\omega(\phi)\,g^{\mu\nu}\partial_\mu \phi\,\partial_\nu \phi - V_{J}(\phi)\bigg],
\label{LJ}
\end{eqnarray}
where a subscript $J$ stands for quantities in the Jordan frame and the reduced Planck mass is defined as $M_p^2 = 1/8\pi G$. Here $\Omega^{2}(\phi)$ is given by
\begin{eqnarray}
\Omega^{2}(\phi) = \frac{M^{2}_{p}+\xi f(\phi)}{M^{2}_{p}},
\label{Ome}
\end{eqnarray}
where $f(\phi)$ is an arbitrary function on the scalar field $\phi$ and $\xi$ is a dimensionless coupling constant. By applying the conformal transformation ${\bar g}_{\mu\nu} = \Omega^{2}(\phi)g_{\mu\nu}$, we eliminate the non-minimal coupling between $f(\phi)$ and the gravitational field. The resulting action in the Einstein frame reads
\begin{eqnarray}
S_E = \int d^4x\sqrt{-g}\,\bigg[ -\frac{1}{2}\,M_p^2\, R + \frac{1}{2}\,g^{\mu\nu}\partial_\mu \psi\,\partial_\nu \psi - U_{E}(\psi)\bigg]\,,
\label{LE}
\end{eqnarray}
where $ U_{E}(\psi)=V_{J}(\phi)/\Omega^{4}(\phi)$ and where a subscript $E$ stands for quantities in the Einstein frame and
\begin{eqnarray}
d\psi^{2} = \bigg[\frac{\omega(\phi)}{\Omega^{2}(\phi)} + 6M^{2}_{p}\frac{\Omega^{'2}(\phi)}{\Omega^{2}(\phi)} \bigg]\,d\phi^{2},\,
\label{Omgg}
\end{eqnarray}
In order to obtain the action in the Einstein frame, we have used the following identities in 4 spacetime dimensions \cite{Fujii2003}:
\begin{eqnarray}
{\bar R} &=& \frac{1}{\Omega^{2}}\bigg[R-6\,g^{\mu\nu}\nabla_{\mu}\nabla_{\nu}\ln\phi - 6\,g^{\mu\nu}(\partial_\mu \ln\phi)(\partial_\nu \ln\phi)\bigg]\,,\nonumber\\
{\bar g}^{\mu\nu} &=& \Omega^{-2}g^{\mu\nu},\,\,\,\sqrt{-{\bar g}}=\Omega^{4}\sqrt{-g}\nonumber\,,
\label{ids}
\end{eqnarray}
where an argument of $\Omega$ is understood, and a bar denotes quantities in the Einstein frame, and we have omitted tildes for convenience. If a conformal factor $\Omega(\phi)$ and a kinetic coupling $\omega(\phi)$ satisfy the condition
\begin{eqnarray}
\omega(\phi)=\frac{M^{2}_{p}}{\xi}\Omega^{'2}(\phi)\,,
\label{cons}
\end{eqnarray}
then there exists an exact relationship between $\phi$ and $\psi$ obtained from Eq.(\ref{Omgg}):
\begin{eqnarray}
\psi=\sqrt{6\alpha}\,M_{p}\ln \Omega,\,\,\,\,\,\,\Omega(\phi)=e^{\sqrt{1/6\alpha}\,\psi/M_{p}}\,.
\label{conso11}
\end{eqnarray}
and
\begin{eqnarray}
V_{J}(\phi)=\Omega^{4}(\phi)U_{E}\big(\sqrt{6\alpha}\,M_{p}\ln \Omega(\phi)\big)\,,
\label{conso}
\end{eqnarray}
where $\alpha=1+(6\xi)^{-1}$. Under the condition (\ref{Omgg}), if we take $V_{J}(\phi)=V_{0}\big(1-\Omega^{2}(\phi)\big)^{2}$, then we get the E-model potential in the Einstein frame
\begin{eqnarray}
U_{E}(\psi)=V_{0}\bigg(1-e^{-2/\sqrt{6\alpha}\,\psi/M_{p}}\bigg)^{2}\,,
\label{conso}
\end{eqnarray}
and in the context of $\alpha$-attractor models we find for small values of $\alpha$ to the leading order of the number of e-folds, $N$:
\begin{eqnarray}
n_{s}=1-\frac{2}{N},\,\,\,\,r=\frac{12\alpha}{N^{2}}\,.
\label{conso}
\end{eqnarray}
We notice that the above result is independent of the function $\Omega(\phi)$. Under the
condition (\ref{Omgg}), if we choose
\begin{eqnarray}
V_{J}(\phi)=V_{0}\Omega^{4}(\phi)\bigg(\frac{1-\Omega^{2}(\phi)}{1+\Omega^{2}(\phi)}\bigg)^{2}\,,
\label{cot}
\end{eqnarray}
then we get the T-model potential in the Einstein frame
\begin{eqnarray}
U_{E}(\psi)=V_{0}\tanh^{2}\bigg(\frac{\psi}{\sqrt{6\alpha}M_{p}}\bigg)\,.
\label{cot2}
\end{eqnarray}
However, the results derived above can be generalized. In so doing, we instead consider $V_{J}(\phi)=V_{0}\Omega^{4}\big(1-\Omega^{-2}(\phi)\big)^{2n}$ and in this case we have
\begin{eqnarray}
U_{E}(\psi)=V_{0}\bigg(1-e^{-2/\sqrt{6\alpha}\,\psi/M_{p}}\bigg)^{2n}\,.
\label{conson}
\end{eqnarray}
In the same manner, we can choose 
\begin{eqnarray}
V_{J}(\phi)=V_{0}\Omega^{4}(\phi)\bigg(\frac{1-\Omega^{2}(\phi)}{1+\Omega^{2}(\phi)}\bigg)^{2n}\,,
\label{cotn}
\end{eqnarray}
consequently to obtain
\begin{eqnarray}
U_{E}(\psi)=V_{0}\tanh^{2n}\bigg(\frac{\psi}{\sqrt{6\alpha}M_{p}}\bigg)\,.
\label{cot2n}
\end{eqnarray}
It is worth noting that a scale of the potential $V_0$ can be determined using the COBE normalization condition. There was a class of inflationary models so called cosmological $\alpha$-attractors has recently received considerable attention \cite{Kallosh:2013yoa,Kallosh:2014rga,Kallosh:2015lwa,Roest:2015qya,Linde:2016uec,Terada:2016nqg,Ueno:2016dim,Odintsov:2016vzz,Akrami:2017cir,Dimopoulos:2017zvq,Pozdeeva:2020shl,Odintsov:2020thl}.

\section{Model setup: Warm inflation}\label{s3}
For the benefit of the reader, we would stress here that in the following we do assume the model present in Ref.\cite{Bastero-Gil:2016qru} for the interactions. Hence after the conformal transformation, we will directly couple the fermions in the Einstein frame Lagrangian (\ref{LE}). Considering the action (\ref{LE}) in the Einstein frame with the flat FLRW line element, the Friedmann equation for warm inflation takes the form
\begin{eqnarray}
H^2 = \frac{1}{3\,M_p^2}\left( \frac{1}{2}\,\dot\psi^2 + U_{E}(\psi) + \rho_r\right)\,,
\end{eqnarray}
where $\dot{\psi}=d\psi/dt$ and $\rho_{r}$ is the energy density of the radiation fluid with the equation of state given by $w_{r}=1/3$. The dynamics of the scalar field ($\psi$) with the dissipative term ($\Gamma$) in the context of warm inflation scenario is governed by the Klein-Gordon equation and is described by the following equation:
\begin{eqnarray}
\ddot\psi + 3H\,\dot\psi + U'_{E}(\psi) = -\Gamma\,\dot\psi\,,
\end{eqnarray}
where $U'_{E}(\psi)=dU_{E}(\psi)/d\psi$. In the warm inflationary scenario, the dissipative coefficient, $\Gamma$, represents the decay of the inflaton field to the radiation. The conservation of the energy-momentum tensor of the radiation fluid is govern by the continuity equation: 
\begin{eqnarray}
\dot\rho_r + 4H\,\rho_r = \Gamma\,\dot\psi^2\,.
\end{eqnarray}
A general form of the dissipative parameter was proposed in Refs.\cite{Berera:1998gx,Berera:2001gs,Zhang:2009ge,BasteroGil:2011xd}. In this study, we consider $\Gamma = C_1\,T$. 
\begin{figure}[!h]	
	\includegraphics[width=7.5cm]{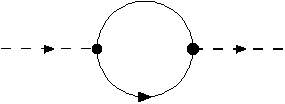}
	\includegraphics[width=7.5cm]{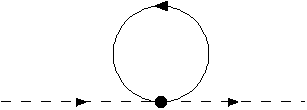}
	\centering
	\caption{Feynman diagrams reproduced from \cite{Bastero-Gil:2016qru} contribute to the inflaton self-energy at one-loop order.}
	\label{stplotE1}
\end{figure}
The dissipation coefficient $\Gamma=\Gamma(\phi,T)$ can be computed from first principles using standard thermal field theory techniques. Let us compute dissipation coefficient in the adiabatic, near thermal equilibrium regime where a
background scalar field (inflaton) is coupled to fermion fields. These interactions are typical of warm inflation microscopic model building. We focus only a regime at high temperature, when the masses of both intermediate and radiation fields are less than the temperature scale. Here we only outline the main
steps of the calculation, see some relevant existing references \cite{Berera:2008ar,Bastero-Gil:2010dgy}.

To compute the inflaton self-energy at one-loop order, the relevant diagrams are shown in Fig.\ref{stplotE1}. However, as mentioned in Ref.\cite{Bastero-Gil:2016qru}, only a left panel yields a nonlocal contribution, with external legs corresponding to different times $t$ and $t'$, whereas a right panel contributes only locally. Therefore, in the present analysis, the left panel mainly contributes to the dissipative term, $\Gamma$. It is given in terms of the retarded inflaton self-energy, $\Sigma_{R}$, in the real-time formalism \cite{Berera:2008ar}. Traditionally, the dissipation coefficient can be computed via \cite{Bartrum:2014fla}
\begin{eqnarray}
\Gamma = \int d^{4}x'\Sigma_{R}(x,x')(t-t'),
\label{radia}
\end{eqnarray}
where the inflaton self-energy due to a fermion loop can be computed using a standard field theory, see for example \cite{Moss:2006gt}. The retarded self-energy can be computed using standard techniques \cite{Kapusta2006}. In high-temperature regime, i.e. $m_{{\hat \chi}_{i}}\ll T$, the leading contributions correspond to on-shell fermions can be determined to yield \cite{Bastero-Gil:2016qru}
\begin{eqnarray}
\Gamma_{i} = 4\frac{g^{2}_{i}}{T}\int \frac{d^{3}p}{(2\pi)^{3}}\frac{m^{2}_{i}}{\Gamma_{{\hat \chi}_{i}}\omega^{2}_{p}}n_{F}(\omega_{p})\big[1-n_{F}(\omega_{p})\big],
\label{gai1}
\end{eqnarray}
where $n_{F}(\omega_{p})$ is the Fermi-Dirac distribution, $\omega_{p}$ is the energy of a state with momentum $p$ given by $\omega_{p}=\sqrt{|{\bf p}|^{2}+m^{2}_{i}}$, and $\Gamma_{{\hat \chi}_{i}}$ is the fermion decay width. Here we consider a dominance channel of additional Yukawa interactions, involving a scalar singlet and chiral fermions in which the Lagrangian was given in Ref.\cite{Bastero-Gil:2016qru}. In this work, we have coupled the fermions after the conformal transformation and then considered the dynamics in the Einstein frame. Therefore, we can deduce the form of the (additional Yukawa) interaction in the Einstein one, and it reads
\begin{eqnarray}
L^{E}_{\psi\hat{\chi}} = h\,\psi\sum_{i=1,2}\big[\bar{\hat{\chi}}_{iL}\hat{\chi}_{\psi R}+\bar{\hat{\chi}}_{\psi L}\hat{\chi}_{i R}\big],
\label{Yu}
\end{eqnarray}
where \lq\lq\,\,$\hat{}$\,\,\rq\rq\, denotes quantities in the Einstein frame, $h$ is the Yukawa coupling, and $\psi$ is the canonically normalized scalar field. Of course, the conformal transformation changes the form of the  original  potential as well as the interaction term. Here we explain in greater
detail what interactions in the Jordan frame transform into this needed interaction in the Einstein frame, e.g. Eq.(\ref{Yu}). We take the Yukawa interaction as an example. This interaction in the Jordam frame can be simply quantified using the following appropriate re-scaling of the fermion \cite{Bezrukov:2010jz}:
\begin{eqnarray}
\hat{\chi}\quad&\rightarrow&\quad \Omega^{-3/2}\chi\,,
\label{res}
\end{eqnarray}
as well as $\sqrt{-{\hat g}}\rightarrow \Omega^{4}\sqrt{-g}$. Substituting these two replacements, we find
\begin{eqnarray}
L^{E}_{\psi\hat{\chi}}\quad\rightarrow\quad L^{J}_{\psi(\phi)\chi} =h\psi(\phi)\Omega(\phi)\sum_{i=1,2}\big[\bar{\chi}_{iL}\chi_{\psi R}+\bar{\chi}_{\psi L}\chi_{i R}\big],
\label{Yu1}
\end{eqnarray}
where $\Omega(\phi)$ is given in Eq.(\ref{Ome}). Therefore, in order to have the needed interaction in the Einstein frame Eq.(\ref{Yu}), we have discovered that the interactions in the Jordan frame are of the form of Eq.(\ref{Yu1}). Comparing the interactions of those two frames, it is reasonable to work in the Einstein frame ones (\ref{Yu}). Considering the interactions (\ref{Yu}), it was found that the on-shell decay width at finite temperature, neglecting the masses of the decay products, yields \cite{Bastero-Gil:2016qru}
\begin{eqnarray}
\Gamma_{\psi_{i}} = \frac{h^{2}}{16\pi}\frac{T^{2}m^{2}_{i}}{\omega^{2}_{p}|{\bf p}|}\Big[F(k_{+}/T,\omega_{p}/T)-F(k_{-}/T,\omega_{p}/T)\Big],
\label{gai12}
\end{eqnarray}
where $k_{\pm}=(\omega_{p}\pm |{\bf p}|)/2$ and $F(x,y)=xy-x^{2}/2+(y-x)\ln(1-e^{-x}/(1+e^{x-y}))+{\rm Li}_{2}(e^{-x})+{\rm Li}_{2}(e^{x-y})$ with ${\rm Li}_{2}(z)$ being the dilogarithm function. The 3-momentum integrals can then be easily computed analytically in different regimes. In particular, for light on-shell modes one typically obtains
$\Gamma_{i}\propto m_{i}\propto T$ yielding $\Gamma = C_{1} T$ for fermionic modes. Here $C_1$ is a function of the coupling $g$ and the Yukawa coupling $h$ determining the decay of the $\psi_{1,2}$ fermions into a light scalar and a light fermion \cite{Bastero-Gil:2018uep}.

Based on a warm inflationary universe in the slow-roll regime, we can re-write the Firedmann equation as well as the equations of motion for the scalaron (inflaton) and the radiation matter as
\begin{eqnarray}
H^2 &\approx& \frac{1}{3 M_p^2}\,U_{E}(\psi)\,,
\label{SR-friedmann}
\\
\dot\phi &\approx& -\frac{U'_{E}(\psi)}{3H(1+Q)}\,,\qquad Q\equiv \frac{\Gamma}{3H}\,,
\label{SR-KG}
\\
\rho_r &\approx& \frac{\Gamma\,\dot\psi}{4H}\,,\qquad \rho_r = C_r\,T^4\,,
\label{SR-rad}
\end{eqnarray}
where $Q$ denotes a dissipative coefficient and $C_r = g_*\,\pi^2/30$. To obtain the above expressions, the following approximations have been used:
\begin{eqnarray}
\rho_r &\ll& \rho_\psi\,,\qquad 
\rho_\psi = \frac12\,\dot\psi^2 + U_{E}\,,
\\
\dot\phi^2 &\ll& U_{E}(\psi)\,,
\\
\ddot\psi &\ll& 3H\left( 1 + Q\right)\dot\psi\,,
\\
\dot\rho_r &\ll& 4H\,\rho_r\,,
\end{eqnarray}
as usually done in the slow-roll scenario. In the strong regime, a model of warm inflation is conducted with the assumption of $Q \gg 1$. More importantly, the temperature can be recast in terms of the scalar field, $\psi$. In the present study, we consider a temperature for $Q\gg 1$ written in the following form:
\begin{eqnarray}
T &=& \left( \frac{U_{E}'^{\,2}\,\psi^{m-1}}{4H\,C_m\,C_r}\right)^{\frac{1}{4+m}}\,,
\label{T-phi-strong}
\end{eqnarray}
with $m$ being any integer. However, in this work we consider $m=1$. In warm inflation, the slow-roll parameters are slightly modified and they take the form
\begin{eqnarray}
\varepsilon &=& \frac{M_p^2}{2}\left( \frac{U'_{E}}{U_{E}}\right)^2\,,\quad \eta = M_p^2\,\frac{U_{E}''}{U_{E}}\,,\quad \beta = M_p^2\left( \frac{U_{E}'\,\Gamma'}{U_{E}\,\Gamma}\right)\,.
\label{SR-parameters}
\end{eqnarray}
Similar to cold inflation, inflationary phase of the universe in warm inflation takes place when the slow-roll parameters satisfy the following conditions:
\begin{eqnarray}
\varepsilon \ll 1 + Q\,,\qquad \eta \ll 1 + Q\,,\qquad \beta \ll 1 + Q\,.\label{sloe}
\end{eqnarray}
Additionally, the number of e-foldings, $N$, in warm inflation gets modified and it can be written for $Q\gg 1$ as 
\begin{eqnarray}
N = \int_{\psi_{\rm end}}^{\psi_N}\frac{Q\,U_{E}}{U'_{E}}\,d\psi\,.
\end{eqnarray}
The power spectrum of the warm inflation was calculated in Refs.\cite{Graham2009,Bastero-Gil:2018uep,Hall:2003zp,Ramos:2013nsa,BasteroGil:2009ec,Taylor:2000ze,DeOliveira:2001he,Visinelli:2016rhn}
and it reads
\begin{eqnarray}
\Delta_{\mathcal{R}} = \left( \frac{H_N^2}{2\pi\dot\psi_N}\right)^2\left( 1 + 2n_N +\left(\frac{T_N}{H_N}\right)\frac{2\sqrt{3}\,\pi\,Q_N}{\sqrt{3+4\pi\,Q_N}}\right)G(Q_N)\,,
\label{spectrum}
\end{eqnarray}
where the subscript $``N"$ is labeled for all quantities estimated at the Hubble horizon crossing and $n = 1/\big( \exp{H/T} - 1 \big)$ is the Bose-Einstein distribution function. More importantly, the function $G(Q_N)$ encodes the coupling between the inflaton and the radiation in the heat bath which leads to a growing mode in the fluctuations of the inflaton field originally studied in Ref.\cite{Graham2009} and consequent implications \cite{BasteroGil:2011xd,BasteroGil:2009ec}. In addition, the scalar spectral index is defined as
\begin{eqnarray}
n_s - 1 = \frac{d\ln \Delta_{\mathcal{R}}}{d\ln k}\Bigg|_{k=k_N} = \frac{d\ln \Delta_{\mathcal{R}}}{dN}\,,
\end{eqnarray}
with $\ln k\equiv a\,H =N$\,. The tensor-to-scalar ratio of the pertubation, $r$, can be calculated via the following formula:
\begin{eqnarray}
r = \frac{\Delta_T}{\Delta_{\mathcal{R}}}\,,
\label{tensor-scalar}
\end{eqnarray}
where $\Delta_T$ is the power spectrum of the tensor perturbation and it takes the same form as the standard (cold) inflation, i.e. $\Delta_T = 2H^2/\pi^2M_p^2 = 2U_{E}(\psi)/3\pi^2 M_p^4$. Therefore the power spectrum in Eq.(\ref{spectrum}) can be recast to yield
\begin{eqnarray}
\Delta_{\mathcal{R}} = \frac{U_{E}(\phi_N)\big(1 + Q_N\big)^2}{24\,\pi^2\,M_p^4\,\varepsilon}\left(1 + 2\,n_N + \left(\frac{T_N}{H_N}\right)\frac{2\sqrt{3}\,\pi\,Q_N}{\sqrt{3+4\pi\,Q_N}}\right)G(Q_N)\,.
\end{eqnarray}
The growth rate of the inflaton field fluctuation from the coupling between the inflaton and the radiation fluid in the thermal bath is repersented by a function $G(Q_N)$ \cite{Graham2009}. For the Higgs-like and plateau-like potentials, the growing mode function was suggested in Ref.\cite{Bastero-Gil:2018uep} and was given by 
\begin{eqnarray}
G_1(Q_N) \simeq 1 + 0.18\,Q_N^{1.4} + 0.01\,Q_N^{1.8}\,,
\label{growing-mode}
\end{eqnarray}
whereas the original growing mode function of the warm little inflation was written as \cite{Bastero-Gil:2016qru}
\begin{eqnarray}
G_2(Q_N) \simeq 1 + 0.335\,Q_N^{1.364} + 0.0185\,Q_N^{2.315}\,.
\label{growing-mode-original}
\end{eqnarray}
The function
$G(Q)$ accounts for the growth of inflaton fluctuations
due to the coupling to radiation and must be determined
numerically. As mentioned in Ref.\cite{Bastero-Gil:2018uep}, this function also exhibits a mild dependence on the form of the scalar potential. For a quartic potential scenario, the function of $G(Q)$ is given in Eq.(\ref{growing-mode-original}); while Eq.(\ref{growing-mode}) for Higgs-like and plateau-like potentials. In the present work, the potentials given in Eq.(\ref{conso}) as well as Eq.(\ref{cot2}) shape as the Higgs-like or plateau-like ones. However, in this work we also consider an another form of $G(Q)$ given in Eq.(\ref{growing-mode-original}). Note that at the thermalized inflaton fluctuation limit, $1+ 2\,n_N \simeq 2\,T_N/H_N $ and $T_N/H_N = 3\,Q_N/C_1$, one can re-write the power spectrum in the following form \cite{Bastero-Gil:2018uep},
\begin{eqnarray}
\Delta_{\mathcal{R}} \simeq \frac{5\,C_1^3}{12\,\pi^4\,g_*\,Q_N^2} \left(1+ \frac{\sqrt{3}\,\pi\,Q_N}{\sqrt{3+4\pi\,Q_N}}\right)G(Q_N)\,
\label{spectrum-grow},
\end{eqnarray}
where $\rho_r/V(\phi) = \varepsilon\,Q/2(1+Q)^2$ has been used to obtain above equation. We note that the above power spectrum in this limit is inexplicitly dependent on the inflaton potential \cite{Bastero-Gil:2018uep}. Then, the tensor-scalar ratio $r$ parameter in this case can be obtained by using Eqs.(\ref{tensor-scalar}) and (\ref{spectrum-grow}). It reads
\begin{eqnarray}
r = \frac{\Delta_T}{\Delta_{\mathcal{R}}} = 16\,\varepsilon\left[ \frac{6\,Q_N^3}{C_1}\left(1+ \frac{\sqrt{3}\,\pi\,Q_N}{\sqrt{3+4\pi\,Q_N}}\right)G(Q_N)\right]^{-1}\,.
\label{tensor-scalar-growE}
\end{eqnarray}
The spectral index of the power spectrum with the growing mode function in Eq.(\ref{growing-mode}) is given by \cite{Bastero-Gil:2018uep,BasteroGil:2009ec,Benetti:2016jhf}
\begin{eqnarray}
n_s &=& 1 + \frac{Q_N}{3+5\,Q_N}\frac{\big(6\,\varepsilon - 2\,\eta \big)}{\Delta_{\mathcal{R}}}\,\frac{d\Delta_{\mathcal{R}}}{dQ_N}\,,
\label{ns-grow}\\
\frac{d\Delta_{\mathcal{R}}}{dQ_N} &=& \frac{5\,C_1^3}{12\,\pi^4\,g_*}\Bigg[\frac{1}{Q_N^2}\left(1+ \frac{\sqrt{3}\,\pi\,Q_N}{\sqrt{3+4\pi\,Q_N}}\right)\frac{dG(Q_N)}{dQ_N}
\nonumber\\
&-& \frac{1}{Q_N^3}\left(2+ \frac{\sqrt{3}\,\pi\,Q_N}{ \sqrt{3+4 \pi\,Q_N}}+\frac{2 \,\sqrt{3}\, \pi^2\,Q_N^2}{(3 + 4\,\pi\, Q_N)^{\frac32}}\right)G(Q_N) \Bigg].\nonumber
\label{diff-spectrum-growEns}
\end{eqnarray}
In this section, we have derived all relevant equations necessary for warm inflation model building. We will compare the results to the observational data in Sec.(\ref{s4}).

\subsection{E model}
In the present analysis, we will consider the warm inflation in the strong regime that the inflaton perturbations are non-trivially affected by the fluctuations of the thermal bath, and the amplitude of the spectrum may get a correction, generically called the ``growing mode”, depending on the value of the dissipative ratio. This was originally conducted by Graham and Moss \cite{Graham2009}. The slow-roll parameters in this case are computed to obtain
\begin{eqnarray}
\varepsilon &=& \frac{4 n^2}{3 \alpha\left(-1+e^{\frac{ \sqrt{\frac{2}{3}} \sqrt{\frac{1}{\alpha}}\psi }{M_p}}\right)^2}\,,\qquad \eta = -\frac{4 n \left(e^{\frac{\sqrt{\frac{2}{3}} \sqrt{\frac{1}{\alpha }}\psi}{M_p}}-2 n\right)}{3\alpha  \left(-1+e^{\frac{\sqrt{\frac{2}{3}} \sqrt{\frac{1}{\alpha}}\psi}{M_p}}\right)^2}\,,
\nonumber\\
\beta &=&\frac{4 n \left(3 n-2e^{\frac{\sqrt{\frac{2}{3}} \sqrt{\frac{1}{\alpha }}\psi}{M_p}}\right)}{15 \alpha \left(-1+e^{\frac{\sqrt{\frac{2}{3}} \sqrt{\frac{1}{\alpha}}\psi}{M_p}}\right)^2}\,.
\label{slow-roll-modelE}
\end{eqnarray}
By using Eqs.(\ref{SR-KG}) and (\ref{T-phi-strong}), we find $Q$ for the strong limit:
\begin{eqnarray}
Q = \frac{\sqrt[5]{2} C_1}{3^{3/5}}\left(\frac{n^2 M_p^4 \left(1-e^{-\frac{\sqrt{\frac{2}{3}} \sqrt{\frac{1}{\alpha }} \phi }{M_p}}\right)^{-2 n}}{\alpha  C_1 V_0 C_r \left(e^{\frac{\sqrt{\frac{2}{3}} \sqrt{\frac{1}{\alpha}} \phi }{M_p}}-1\right)^2}\right)^{1/5}\,.
\label{Q-strongE}
\end{eqnarray}
When inflation ends, one finds from Eq.(\ref{sloe}) using a condition $\varepsilon_{\rm end} \approx Q_{\rm end}$:
\begin{eqnarray}
\frac{4 n^2}{3 \alpha\left(-1+e^{\frac{\sqrt{\frac{2}{3}} \sqrt{\frac{1}{\alpha }}\psi_{\rm end} }{M_p}}\right)^2} \approx \frac{\sqrt[5]{2} C_1}{3^{3/5}}\left(\frac{n^2 M_p^4 \left(1-e^{-\frac{\sqrt{\frac{2}{3}} \sqrt{\frac{1}{\alpha }} \phi }{M_p}}\right)^{-2 n}}{\alpha  C_1 V_0 C_r \left(e^{\frac{\sqrt{\frac{2}{3}} \sqrt{\frac{1}{\alpha}} \phi }{M_p}}-1\right)^2}\right)^{1/5}\,,\label{endE}
\end{eqnarray}
Apparently, the above equation can not be analytically solved to obtain exact solutions. However, certain approximate solutions during inflation can be obtained by invoking a large field approximation. To this end, we first assume $\sqrt{\frac{2}{3}} \sqrt{\frac{1}{\alpha }}\psi_{\rm end}/M_p \gg 1$ and we can solve Eq.(\ref{endT}) to obtain a value of the inflaton field at the end of inflation to yield
\begin{eqnarray}
\psi_{\rm end} \approx \sqrt{\frac{3\alpha}{2}} M_p \log \left(\frac{1.657 n \sqrt[8]{V_0} \sqrt[8]{C_r}}{\sqrt{\alpha } \sqrt{C_1} \sqrt{M_p}}\right),
\end{eqnarray}
where a large field approximation has been done by assuming $e^{\sqrt{\frac{2}{3}} \sqrt{\frac{1}{\alpha }}\psi_{\rm end}/M_p}\pm1 \approx e^{\sqrt{\frac{2}{3}} \sqrt{\frac{1}{\alpha }}\psi_{\rm end}/M_p}$. Moreover, the inflaton field at the Hubble horizon crossing in the strong regime, $\phi_{N}$, can be determined to obtain
\begin{eqnarray}
N &=& \frac{1}{M_p^2}\int_{\phi_{\rm end}}^{\phi_N}\frac{Q\,U_{E}}{U'_{E}}\,d\psi
\nonumber\\
&=& \frac{\sqrt{\frac{1}{\alpha }} n V_0 \left(1-e^{-\frac{\sqrt{\frac{2}{3}} \sqrt{\frac{1}{\alpha }} \psi }{M_p}}\right)^{2 n}}{2\ 2^{3/10} \sqrt[10]{3} C_r M_p \left(e^{\frac{\sqrt{\frac{2}{3}} \sqrt{\frac{1}{\alpha }} \phi }{M_p}}-1\right) \left(\frac{n^2 M_p^2 \left(\frac{V_0 \left(1-e^{-\frac{\sqrt{\frac{2}{3}} \sqrt{\frac{1}{\alpha }} \psi }{M_p}}\right){}^{2 n}}{M_p^2}\right){}^{3/2}}{\alpha  C_1 C_r \left(e^{\frac{\sqrt{\frac{2}{3}} \sqrt{\frac{1}{\alpha }} \psi }{M_p}}-1\right)^2}\right)^{4/5}}d\psi
\nonumber\\
&\approx& \frac{5 n V_0}{12\ 2^{3/10} \sqrt[10]{3} \gamma  \sqrt{\alpha}C_r}\left(\frac{e^{-\frac{\sqrt{\frac{2}{3}} \sqrt{\frac{1}{\alpha }}\psi}{M_p}}}{\left(\frac{n^2 V_0 \sqrt{\frac{V_0}{M_p^2}} e^{-\frac{2\sqrt{\frac{2}{3}} \sqrt{\frac{1}{\alpha }}\psi}{M_p}}}{\alpha  C_1 C_r}\right)^{4/5}}\right)^{\psi_{N}}_{\psi_{\rm end}}\,,
\label{N-strongE}
\end{eqnarray}
Assuming $\psi_{N}\gg \psi_{\rm end}$, the above relation becomes
\begin{eqnarray}
N\approx \frac{5 M_p e^{\frac{\sqrt{6} \sqrt{\frac{1}{\alpha }} \psi_{N} }{5 M_p}} \sqrt[5]{\frac{C_1^4}{\left(\frac{1}{\alpha }\right)^{3/2} n^3 V_0 C_r M_p}}}{2\ 2^{4/5} 3^{3/5} \sqrt{\frac{1}{\alpha}}}\,.
\label{expphi-N-strong}
\end{eqnarray}
We then obtain
\begin{eqnarray}
\psi_N=\frac{\sqrt{\alpha} M_p}{\sqrt{6}}\log\left(\frac{4.424 n^3 N^5 V_0 C_r}{\alpha^4 C_1^4 M_p^4}\right)\,.
\label{expphi-N-strongN}
\end{eqnarray}
As done above, we therefore can re-write the slow-roll parameters in terms of the number of {\it e}-foldings, $N$, by using large field approximation in the strong $Q$ limit and then we find
\begin{eqnarray}
\varepsilon \approx \frac{204.73 \alpha  C_1^{8/5}}{N^2 C_r^{2/5}}\,,\qquad \eta = -\frac{16.52 C_1^{4/5}}{N \sqrt[5]{C_r}}\,,\quad \beta = -\frac{6.61 C_1^{4/5}}{N \sqrt[5]{C_r}}\,.
\label{slow-roll-modelEN}
\end{eqnarray}
It is noticed that for a large field approximation the results given in Eq.(\ref{slow-roll-modelEN}) do not depend on an integer, $n$.
\subsection{T model}
The slow-roll parameters for this model can be determined by following computations given in the previous subsection. For this model, we find
\begin{eqnarray}
\varepsilon &=& \frac{4 n^2 \csch^2\left(\frac{\sqrt{\frac{2}{3}} \psi }{\sqrt{\alpha } M_{p}}\right)}{3 \alpha }\,,\qquad \eta = -\frac{4 n \csch^2\left(\frac{\sqrt{\frac{2}{3}} \psi }{\sqrt{\alpha } M_p}\right) \left(\cosh \left(\frac{\sqrt{\frac{2}{3}} \psi }{\sqrt{\alpha } M_p}\right)-2 n\right)}{3 \alpha }\,,
\nonumber\\
\beta &=& \frac{4 n \csch^2\left(\frac{\sqrt{\frac{2}{3}} \psi }{\sqrt{\alpha } M_p}\right) \left(3 n-2 \cosh \left(\frac{\sqrt{\frac{2}{3}} \psi }{\sqrt{\alpha } M_p}\right)\right)}{15 \alpha }\,,
\label{slow-roll-modelT}
\end{eqnarray}
A dissipative coefficient $Q$ in this case can be computed to yield
\begin{eqnarray}
Q = \frac{\sqrt[5]{2} C_1}{3^{3/5}}\left(\frac{n^2 M_p^4 \csch^2\left(\frac{\sqrt{\frac{2}{3}} \psi }{\sqrt{\alpha } M_p}\right) \tanh ^{-2 n}\left(\frac{\psi }{\sqrt{6} \sqrt{\alpha } M_p}\right)}{\alpha  C_1 C_{r} V_0}\right)^{1/5}\,.
\label{Q-strong}
\end{eqnarray}
One deduce from Eq.(\ref{sloe}) when inflation ends with a condition $\varepsilon_{\rm end} \approx Q_{\rm end}$ and find
\begin{eqnarray}
\frac{4 n^2 \csch^2\left(\frac{\sqrt{\frac{2}{3}} \psi }{\sqrt{\alpha } M_p}\right)}{3 \alpha } \approx \frac{\sqrt[5]{2} C_1}{3^{3/5}}\left(\frac{n^2 M_p^4 \csch^2\left(\frac{\sqrt{\frac{2}{3}} \psi }{\sqrt{\alpha } M_p}\right) \tanh^{-2 n}\left(\frac{\psi }{\sqrt{6} \sqrt{\alpha } M_p}\right)}{\alpha  C_1 C_{r} V_0}\right)^{1/5}\,.\label{endT}
\end{eqnarray}
Apparently, the above equation can not be analytically solved to obtain exact solutions. However, certain approximations during inflation can be assumed and the solutions can be computed by using the large field approximation to obtain
\begin{eqnarray}
\sinh\left(\frac{\sqrt{\frac{2}{3}} \psi }{\sqrt{\alpha } M_p}\right) &\approx& e^{\frac{\sqrt{\frac{2}{3}} \psi }{\sqrt{\alpha } M_p}}/2\,,\,\,\,\sinh\left(\frac{\sqrt{\frac{2}{3}} \psi }{\sqrt{\alpha } M_p}\right)=\frac{1}{\csch\left(\frac{\sqrt{\frac{2}{3}} \psi }{\sqrt{\alpha } M_p}\right)}\,,\nonumber\\ \tanh\left(\frac{\sqrt{\frac{1}{6}} \psi }{\sqrt{\alpha } M_p}\right)\nonumber&\approx& 1\,,
\end{eqnarray}
Using the above approximations, we can solve Eq.(\ref{endT}) to obtain the value of the inflaton field at the end of inflation:
\begin{eqnarray}
\psi_{\rm end} \approx \sqrt{\frac{3\alpha}{8}}M_p \log \left(\frac{16 \sqrt[4]{2} n^2 \sqrt[4]{V_0} \sqrt[4]{C_r}}{\sqrt{3} \alpha  C_1 M_p}\right)\,.
\end{eqnarray}
Moreover, the inflaton field at the Hubble horizon crossing in the strong regime, $\phi_{N}$, can be determined to obtain
\begin{eqnarray}
N &=& \frac{1}{M_p^2}\int_{\psi_{\rm end}}^{\psi_N}\frac{Q\,U_{E}}{U'_{E}}\,d\psi
\nonumber\\
&=& \int_{\psi_{\rm end}}^{\psi_N}\frac{n V_0 \csch\left(\frac{\sqrt{\frac{2}{3}} \psi }{\sqrt{\alpha } M_p}\right) \tanh^{2 n}\left(\frac{\psi }{\sqrt{6} \sqrt{\alpha } M_p}\right)}{4 \sqrt[10]{6} \sqrt{\alpha } C_r M_p \left(\frac{n^2 M_p^2 \left(\frac{V_0 \tanh ^{2 n}\left(\frac{\psi }{\sqrt{6} \sqrt{\alpha } M_p}\right)}{M_p^2}\right){}^{3/2}}{\alpha  C_1 C_r \left(\cosh \left(\frac{2 \sqrt{\frac{2}{3}} \psi }{\sqrt{\alpha} M_p}\right)-1\right)}\right){}^{4/5}}d\psi
\nonumber\\
&\approx& \frac{5 n V_0}{4\ 2^{2/5} 3^{3/5} C_r}\left(\frac{e^{-\frac{\sqrt{\frac{2}{3}} \psi _N}{\sqrt{\alpha } M_p}}}{\left(\frac{n^2 V_0 \sqrt{\frac{V_0}{M_p^2}} e^{-\frac{2 \sqrt{\frac{2}{3}} \psi _N}{\sqrt{\alpha } M_p}}}{\alpha  C_1 C_r}\right){}^{4/5}}-\frac{e^{-\frac{\sqrt{\frac{2}{3}} \psi _{\text{end}}}{\sqrt{\alpha } M_p}}}{\left(\frac{n^2 V_0 \sqrt{\frac{V_0}{M_p^2}} e^{-\frac{2 \sqrt{\frac{2}{3}} \psi _{\text{end}}}{\sqrt{\alpha } M_p}}}{\alpha  C_1 C_r}\right){}^{4/5}}\right)\,,
\label{N-strong}
\end{eqnarray}
where we have used the approximated functions given above to obtain the line of Eq.(\ref{N-strong}). Assuming $\psi_{N}\gg \psi_{\rm end}$ the above relation becomes
\begin{eqnarray}
N=\frac{5 n V_0 e^{-\frac{\sqrt{\frac{2}{3}} \psi _N}{\sqrt{\alpha } M_p}}}{4\ 2^{2/5} 3^{3/5} C_r \left(\frac{n^2 V_0 \sqrt{\frac{V_0}{M_p^2}} e^{-\frac{2 \sqrt{\frac{2}{3}} \psi _N}{\sqrt{\alpha} M_p}}}{\alpha  C_1 C_r}\right)^{4/5}}\,,
\label{expphi-N-strongT}
\end{eqnarray}
Thus we obtain
\begin{eqnarray}
\psi_N=\sqrt{\frac{3 \alpha }{2}} M_p \log \left(\frac{48 n N^{5/3} \sqrt[3]{V_0} \sqrt[3]{C_r}}{5\ 5^{2/3} \alpha ^{4/3} C_1^{4/3} M_p^{4/3}}\right)\,,
\label{expphi-N-strongTT}
\end{eqnarray}
As is was done above, we therefore can re-write the slow-roll parameters in terms of the number of {\it e}-foldings, $N$, by using the large field approximation in the strong $Q$ limit to yield
\begin{eqnarray}
\varepsilon = \frac{355.45 \alpha  C_1^{8/5}}{N^2 C_r^{2/5}}\,,\qquad \eta = -\frac{21.8 C_1^{4/5}}{N \sqrt[5]{C_r}}\,,\quad \beta = -\frac{8.71 C_1^{4/5}}{N \sqrt[5]{C_r}}\,,
\label{slow-roll-modelTN}
\end{eqnarray}

Similar to the previous model, we also noticed that for a large field approximation the results given in Eq.(\ref{slow-roll-modelTN}) do not depend on an integer, $n$.

\section{Confrontation with the Planck 2018 data}\label{s4}
In this section, we constrain the inflation potentials using the COBE normalization condition \cite{Bezrukov:2008ut}. This can be used to fix the parameters of the models in the present analysis. From Planck 2018 data, the inflaton potential must be normalized by the slow-roll parameter, $\epsilon$ and satisfied the following relation at the horizon crossing $\psi=\psi_{N}$ in order to produce the observed amplitude of the cosmological density perturbation ($A_{s}$):
\begin{eqnarray}
\frac{U_{E}(\psi_N)}{\epsilon(\psi_N)} \simeq (0.0276\,M_{p})^{4}\,.
\label{Cobe-strong}
\end{eqnarray}
The above constraints is useful in constraining the energy scale of the potential parametrized by $V_{0}$. Additionally, in our analysis below, we compute the inflationary observables $n_{s}$ and $r$, and then compare the results with the Planck 2018 observational constraints.
\subsection{E model}
We start by following Eq.(\ref{Cobe-strong}) and substituting $U(\psi)$ and $\epsilon(\psi)$ for $\psi$ evaluated at the horizon crossing, $\psi_{N}$, and we find for the E model the potential scale $V_{0}$:
\begin{eqnarray}
V_{0} \approx \frac{1.188\times 10^{-4} \alpha  C_1^{8/5} M_p^4}{N^2 C_r^{2/5}}\,.
\label{CobeE}
\end{eqnarray}
It is noticed that $V_{0}$ depend on $\alpha,\,C_{r},\,C_{1}$ and the number of e-foldings $N$. Substituting $\alpha,\,C_{r},\,C_{1},\,N$, we can easily determine the potential scale of the model. Using Eq.(\ref{tensor-scalar-growE}) and Eq.(\ref{diff-spectrum-growEns}), we find for $G_1$
\begin{eqnarray}
r_{1}&\approx&\frac{0.560447 \alpha  \left(\frac{C_1^4}{C_r}\right)^{2/5} C_r^{3/5}}{C_1^{7/5} N^2 \left(BA\right)}\,,\label{rG1E}\\
n_{1s}&\approx& 1+\frac{974.13 \left(\frac{C_1^4}{C_r}\right)^{3/5} \left(1228.38 \alpha  C_1^{8/5}+33.04 C_1^{4/5} N \sqrt[5]{C_r}\right)}{A B N^2 \left(49.57 \sqrt[5]{\frac{C_1^4}{C_r}}+3\right) C_r^{2/5}}\times\nonumber\\&&\times\left(\frac{0.0102 B D}{\left(\frac{C_1^4}{C_r}\right)^{2/5}}-\frac{0.00103 A (B+C+2)}{\left(\frac{C_1^4}{C_r}\right)^{3/5}}\right)\,,
\end{eqnarray}
and for $G2$
\begin{eqnarray}
r_{2}&\approx&\frac{0.560447 \alpha\left(\frac{C_1^4}{C_r}\right)^{2/5} C_r^{3/5}}{C_1^{7/5} N^2 B E}\,,\\
n_{2s}&\approx&1+\frac{974.13 \left(\frac{C_1^4}{C_r}\right){}^{3/5} \left(\frac{0.010176 B F}{\left(\frac{C_1^4}{C_r}\right){}^{2/5}}-\frac{0.00103 e (B+C+2)}{\left(\frac{C_1^4}{C_r}\right)^{3/5}}\right)}{E B N^2 \left(49.565 \sqrt[5]{\frac{C_1^4}{C_r}}+3\right) C_r^{2/5}}\times\nonumber\\&&\times\left(1228.38 \alpha  C_1^{8/5}+33.04 C_1^{4/5} N \sqrt[5]{C_r}\right)\,,\label{rEG2}
\end{eqnarray}
where we have defined new parameters
\begin{eqnarray}
A&\equiv&1+0.62 \left(\frac{C_1^4}{C_r}\right)^{0.36}+4.47 \left(\frac{C_1^4}{C_r}\right)^{0.28}\,,\nonumber\\
B&\equiv&1+\frac{53.94 \sqrt[5]{\frac{C_1^4}{C_r}}}{\sqrt{124.57 \sqrt[5]{\frac{C_1^4}{C_r}}+3}}\,,\,\,
C\equiv\frac{3359.7 \left(\frac{C_1^4}{C_r}\right)^{2/5}}{\left(124.57 \sqrt[5]{\frac{C_1^4}{C_r}}+3\right)^{3/2}}\,,\nonumber\\
D&\equiv&0.113 \left(\frac{C_1^4}{C_r}\right)^{0.16}+0.631 \left(\frac{C_1^4}{C_r}\right)^{0.08}\,,\nonumber\\
E&\equiv&1+3.74 \left(\frac{C_1^4}{C_r}\right)^{0.463}+7.65 \left(\frac{C_1^4}{C_r}\right)^{0.2728}\,,\nonumber\\
D&\equiv&0.87 \left(\frac{C_1^4}{C_r}\right)^{0.263}+1.05\left(\frac{C_1^4}{C_r}\right)^{0.0728}\,.\nonumber
\end{eqnarray}
\begin{figure}[!h]	
	\includegraphics[width=7.5cm]{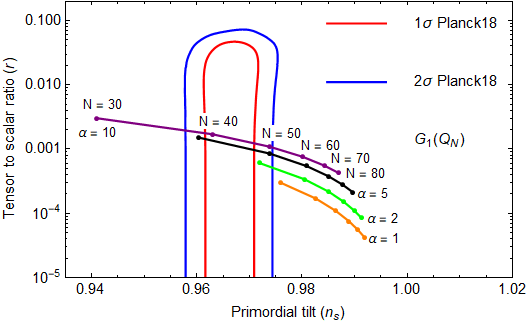}
	\includegraphics[width=7.5cm]{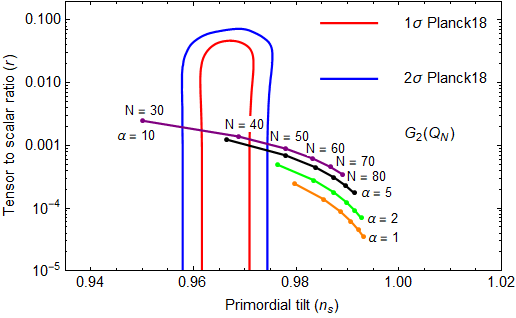}
	\centering
	\caption{We compare the theoretical predictions of $(r,\,n_s)$ in the strong limit $Q>>1$ for the E model. We consider two forms of $G(Q_N)$ given in Eq.(\ref{growing-mode}) for $G_{1}(Q_N)$ (left panel) \& (\ref{growing-mode-original}) for $G_{2}(Q_N)$ (right panel) with $C_{1}=0.28,\,C_{r}=70$. We consider theoretical predictions of $(r,\,n_s)$ for different values
of $N$ and $\alpha$ with Planck’18 results for TT, TE, EE, +lowE+lensing+BK15+BAO.}
	\label{stplotE}
\end{figure}
We have used $\varepsilon$ and $\eta$ given in Eq.(\ref{slow-roll-modelEN}) in order to obtain the results present in Eq.(\ref{rG1E}-\ref{rEG2}). 

To compare the results with the observations, we plot the ($n_{s}-r$) graph for the two models of $G_{i}(Q)$ with $i=1,2$ along with the observational constraints from Planck 2018 data in Fig.(\ref{stplotE}). Left panel, we show our results obtained by considering $G_{1}(Q)$ and find that the large values of $\alpha\gg 1$ is needed in order to have the predictions of $n_s$ and $r$ in agreement with the data. In other words, the derived $n_s$ and $r$ are consistent with the Planck observations only in the weak coupling regime $\xi\ll 1$ since $\alpha=1+1/(6\xi)$. Likewise, for the right panel, we consider $G_{2}(Q)$ and find that the large values of $\alpha\gg 1$ is required in order to have the derived $n_s$ and $r$ consistent with the Planck 2018 observations.  

We find for $G_{1}(Q)$ using $C_{r}=70,\,C_{1}=0.28$ and $N=55$ of the E model $n_{s}=0.965$ and $r=1.87\times 10^{-3}$ for $\alpha=20.75$ implying $\xi=0.00845$, whilst $G_{2}(Q)$ using $C_{r}=70,\,C_{1}=0.28$ and $N=55$ of the E model $n_{s}=0.965$ and $r=1.94\times 10^{-3}$ for $\alpha=26.32$ implying $\xi=0.00658$. Interestingly, we discover that the energy scale of the potential $V^{1/4}_{0}\approx 1.18\times 10^{-2}\,M_{p}\sim 2.89\times 10^{16}$\,{\rm GeV} of the E model with $M_{p}$ being the reduced Planck mass.

\subsection{T model}
For the T model, we also consider a constraint given in Eq.(\ref{Cobe-strong}) and substitute $U(\psi)$ and $\epsilon(\psi)$ for $\psi$ evaluated at the horizon crossing, $\psi_{N}$, and we obtain the potential scale $V_{0}$:
\begin{eqnarray}
V_{0}\approx 5.17\times 10^{-5}\frac{\alpha\ C_1^{8/5}M_{p}^4}{N^2 C_r^{2/5}}\,,
\end{eqnarray}
We see that $V_{0}$ also depend on $\alpha,\,C_{r},\,C_{1}$ and the number of e-foldings $N$ similar to that of the E model. Using $\alpha,\,C_{r},\,C_{1},\,N$, we can determine the potential scale of the model to yield for $G1$
\begin{eqnarray}
r_{1}&\approx&\frac{0.423375 \alpha  \left(\frac{C_1^4}{C_r}\right){}^{2/5} C_r^{3/5}}{{\tilde A} {\tilde B} C_1^{7/5} N^2}\\
n_{1s}&\approx& 1+\frac{2238.84 \left(\frac{C_1^4}{C_r}\right){}^{3/5} \left(2132.7 \alpha  C_1^{8/5}+43.6 C_1^{4/5} N \sqrt[5]{C_r}\right)}{{\tilde A} {\tilde B} N^2 \left(65.41 \sqrt[5]{\frac{C_1^4}{C_r}}+3\right) C_r^{2/5}}\times\nonumber\\&&\times\left(\frac{0.00584321 {\tilde A} {\tilde D}}{\left(\frac{C_1^4}{C_r}\right)^{2/5}}-\frac{0.000446661 {\tilde B} ({\tilde A}+{\tilde B}+2)}{\left(\frac{C_1^4}{C_r}\right)^{3/5}}\right)\,,
\end{eqnarray}
and find for $G2$
\begin{eqnarray}
r_{2}&\approx&\frac{0.423375 \alpha  \left(\frac{C_1^4}{C_r}\right)^{2/5} C_r^{3/5}}{C_1^{7/5} N^2 {\tilde A} {\tilde E}}\,,\\
n_{2s}&\approx&1+\frac{358.362 \left(\frac{C_1^4}{C_r}\right)^{3/5} \left(1228.38 \alpha  C_1^{8/5}+33.04 C_1^{4/5} N \sqrt[5]{C_r}\right)}{B N^2 \left(49.565 \sqrt[5]{\frac{C_1^4}{C_r}}+3\right) C_r^{2/5}}\times\nonumber\\&&\times\left(\frac{0.010176 B F}{\left(\frac{C_1^4}{C_r}\right)^{2/5}}-\frac{0.00279983 (B+C+2)}{\left(\frac{C_1^4}{C_r}\right)^{3/5}}\right)\,,
\end{eqnarray}
where we have defined new parameters
\begin{eqnarray}
{\tilde A}&\equiv&1+\frac{71.1844 \sqrt[5]{\frac{C_1^4}{C_r}}}{\sqrt{164.393 \sqrt[5]{\frac{C_1^4}{C_r}}+3}}\,,\nonumber\\
{\tilde B}&\equiv&1+1.02332 \left(\frac{C_1^4}{C_r}\right)^{0.36}+6.58592 \left(\frac{C_1^4}{C_r}\right)^{0.28}\,,\nonumber\\
{\tilde C}&\equiv&\frac{5851.12 \left(\frac{\text{C1}^4}{C_{r}}\right)^{2/5}}{\left(164.393 \sqrt[5]{\frac{C_{1}^4}{C_{r}}}+3\right)^{3/2}}\,,\nonumber\\
{\tilde D}&\equiv&0.140803 \left(\frac{C_{1}^4}{C_{r}}\right)^{0.16}+0.704807 \left(\frac{C_{1}^4}{C_{r}}\right)^{0.08}\,,\nonumber\\
{\tilde E}&\equiv&1+7.1166 \left(\frac{C_1^4}{C_r}\right)^{0.463}+11.1735 \left(\frac{C_1^4}{C_r}\right)^{0.2728}\,.\nonumber\\
{\tilde F}&\equiv &1.25936 \left(\frac{C_1^4}{C_r}\right)^{0.263}+1.16501 \left(\frac{C_1^4}{C_r}\right)^{0.0728}\,.\nonumber
\end{eqnarray}
\begin{figure}[!h]	
	\includegraphics[width=7.5cm]{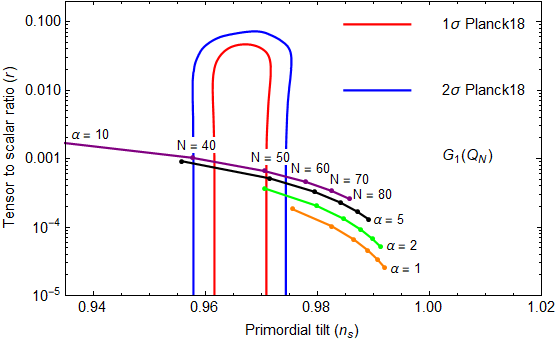}
	\includegraphics[width=7.5cm]{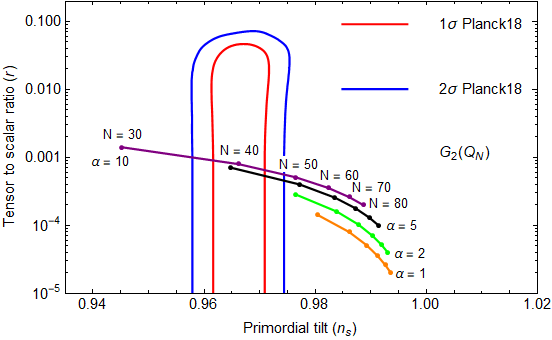}
	\centering
	\caption{We compare the theoretical predictions of $(r,\,n_s)$ in the strong limit $Q>>1$ for the T model. We consider two forms of $G(Q_N)$ given in Eq.(\ref{growing-mode}) for $G_{1}(Q_N)$ (left panel) \& (\ref{growing-mode-original}) for $G_{2}(Q_N)$ (right panel) with $C_{1}=0.28,\,C_{r}=70$. We consider theoretical predictions of $(r,\,n_s)$ for different values
of $N$ and $\alpha$ with Planck’18 results for TT, TE, EE, +lowE+lensing+BK15+BAO.}
	\label{stplotT}
\end{figure}

We plot the derived $n_{s}$ and $r$ for the two models of $G_{i}(Q)$ with $i=1,2$ along with the observational constraints from Planck 2018 data in Fig.(\ref{stplotT}). Left panel, we show our results obtained by considering $G_{1}(Q)$ and find that the Planck data prefers large values of $\alpha\gg 1$. In other words, the derived $n_s$ and $r$ are consistent with the Planck observations only in the weak coupling regime $\xi\ll 1$ since $\alpha=1+1/(6\xi)$. Likewise, for the right panel, we consider $G_{2}(Q)$ and find that the large values of $\alpha\gg 1$ is required in order to have the derived $n_s$ and $r$ consistent with the Planck 2018 observations.  

We find for $G_{1}(Q)$ using $C_{r}=70,\,C_{1}=0.28$ and $N=55$ of the T model $n_{s}=0.965$ and $r=9.06\times 10^{-4}$ for $\alpha=16.62$ implying $\xi=0.017$, whilst $G_{2}(Q)$ using $C_{r}=70,\,C_{1}=0.28$ and $N=55$ of the T model $n_{s}=0.965$ and $r=9.54\times 10^{-4}$ for $\alpha=22.65$ implying $\xi=0.0077$. Interestingly, we discover that the energy scale of the potential  $V^{1/4}_{0}\approx 9.62\times 10^{-3}\,M_{p}\sim 2.35\times 10^{16}$\,{\rm GeV} of the T model with $M_{p}$ being the reduced Planck mass.

\section{Conclusion}

In summary, we have investigated warm inflationary models in the context of a general scalar-tensor theory of gravity which is coupled to radiation through a dissipation term. We presented detailed derivations of the potentials: exponential and hyperbolic tangent forms. We have derived relevant dynamical equations in the non-minimal coupling warm inflation under the slow-roll approximation. 

In this work, we have particularly considered a dissipation parameter of the form $\Gamma = C_{1}T$ with $C_1$ being a coupling parameter and have focused only on the strong regime of which the interaction between inflaton and radiation fluid has been taken into account. To test the results, we have computed inflationary observables and have constrained the parameters of our model using current Planck 2018 data. Additionally, the potential scale $V_{0}$ of the models were constrained using the COBE normalization condition. Interestingly, we have found that a weak coupling limit $\xi\ll 1$ is needed in order to have the derived $n_s$ and $r$ consistent with the Planck 2018 observational constraints. Notice that small values of the non-minimal coupling $\xi \gg 0.003$ were reported in the previous works on cold inflation, see for example \cite{Bezrukov:2013fca,Boubekeur:2015xza}.

\section*{Acknowledgements}
P. Channuie acknowledged the Mid-Career Research Grant 2020 from National Research Council of Thailand (No.NRCT5-RSA63019-03) and is partially supported by the National Science, Research and Innovation Fund (SRF) with grant No. P2565B202.

\end{document}